# Model-based image adjustment for a successful pansharpening

**Gintautas Palubinskas**

Remote Sensing Technology Institute,
German Aerospace Center (DLR),
Oberpfaffenhofen, 82234 Wessling, Germany,
E-Mail: Gintautas.Palubinskas@dlr.de

**Abstract:** A new model-based image adjustment for the enhancement of multi-resolution image fusion or pansharpening is proposed. Such image adjustment is needed for most pansharpening methods using panchromatic band and/or intensity image (calculated as a weighted sum of multispectral bands) as an input. Due various reasons, e.g. calibration inaccuracies, usage of different sensors, input images for pansharpening: low resolution multispectral image or more precisely the calculated intensity image and high resolution panchromatic image may differ in values of their physical properties, e.g. radiances or reflectances depending on the processing level. But the same objects/classes in both images should exhibit similar values or more generally similar statistics. Similarity definition will depend on a particular application. For a successful fusion of data from two sensors the energy balance between radiances/reflectances of both sensors should hold. A virtual band is introduced to compensate for total energy disbalance in different sensors. Its estimation consists of several steps: first, weights for individual spectral bands are estimated in a low resolution scale, where both multispectral and panchromatic images (low pass filtered version) are available, then, the estimated virtual band is up-sampled to a high scale and, finally, high resolution panchromatic band is corrected by subtracting virtual band. This corrected panchromatic band is used instead of original panchromatic image in the following pansharpening. It is shown, for example, that the performance quality of component substitution based methods can be increased significantly.

## 1. Introduction

Pansharpening aims to restore/estimate a multispectral image (MS) in a high resolution space from two given inputs: low resolution multispectral image and high resolution panchromatic image (Pan). Multi-resolution image fusion is not limited only to multispectral and panchromatic image pairs. For example, in hyper-sharpening, a high resolution image is a multispectral image, and a low resolution image is a hyperspectral image. A large number of algorithms and methods to solve this problem were introduced during the last three decades, e.g. [1-6] with different method classifications or groupings have been proposed.

Recently it has been shown that most known and popular methods can be divided into two large groups [7-11]. The first group of methods are named component substitution (CS) methods and are based on a linear spectral transformation, e.g. Intensity-Hue-Saturation (IHS), Principal Component Analysis (PCA), and Gram-Schmidt orthogonalization (GS). Methods of the second group use a spatial frequency decomposition usually performed by means of high pass filtering (HPF), e.g. boxcar filter in signal domain, filtering in Fourier domain or Multi-Resolution Analysis (MRA) using wavelet transform. Additionally, it should be mentioned, that presently a third group of methods spreading quite rapidly in remote sensing community. These are so-called model-based optimization methods based on the minimization of model error residuals, e.g. using Bayesian [12], unmixing [13-15] or sparse representation (SR) based methods [16]. Sometimes these methods are also called variational optimization (VO) based methods, e.g. see review in [11].

Two main pansharpening improvement techniques such as panchromatic image histogram matching [17] and inclusion of image formation/acquisition model into pansharpening method [2, 12, 18-22] have been widely spread during the recent decade.

This paper presents a new model-based image adjustment for the enhancement of multi-resolution image fusion or pansharpening. A virtual band is introduced to compensate for total energy disbalance in different sensors and is used to correct panchromatic image before fusion. Additionally, weights estimated during this correction can be used for the intensity image calculation if appropriate. The paper is organized as follows. In

Sect. 2. the methodology is presented. In Sect. 3 data, experiments and results are described. Finally, the paper ends with discussion, conclusion and reference sections.

## 2. Methodology

In the following sub-sections definitions of images and methodology used in this paper are presented.

*2.1. Definitions*

Below definitions/notations for images and their pixels are introduced which are used in this paper.

2.1.1. Low resolution images

$S_{lr} = \{s_{lr,k,i}, \ k = 1, \ldots, K, \ i = 1, \ldots, I\}$ – low resolution multispectral image matrix (also MS image or MS bands), $k$ – band index, $K$ - number of bands, $i$ – pixel index (for simplicity displayed as a one-dimensional index though in reality it is a two-dimensional index), $I$- number of pixels. It is an input image for pansharpening.

$I_{lr} = \{i_{lr,i}, \ i = 1, \ldots, I\}$ - intensity image calculated from multispectral image $S_{lr}$ using weights $W = \{w_k\}$ based on sensor spectral response functions (usually available from data provider)

$$i_{lr,i} = \sum_1^K w_k \cdot s_{lr,k,i} \tag{1}$$

or in matrix form

$$I_{lr} = W \cdot S_{lr} . \tag{2}$$

$P_{lr} = \{p_{lr,i}, \ i = 1, \ldots, I\}$ – low resolution panchromatic image (Pan image) calculated from high resolution panchromatic image $P_{hr}$ (for definition see the following sub-section) using a low pass filter $F = \{f_j\}$ designed to account resolution change from high to low resolution

$$p_{lr,i} = \sum_{j \in F} f_j \cdot p_{hr,j} \tag{3}$$

where $f_j$ – low pass filter values, index j runs all indices inside a low pass filter $F$. $F$ can include decimation or sub-sampling operator. In matrix form

$$P_{lr} = F \cdot P_{hr} . \tag{4}$$

2.1.2. High resolution images

$S_{hr} = \{s_{hr,k,j}, \ k = 1, \ldots, K, \ j = 1, \ldots, J\}$ – high resolution multispectral image, $k$ – band index, $j$ – pixel index, $J$- number of pixels. It is the aim/result of pansharpening or an output of pansharpening.

$I_{hr} = \{i_{hr,j}, \ j = 1, \ldots, J\}$ - intensity image calculated from multispectral image $S_{hr}$ using weights $W$ based on spectral response functions

$$i_{hr,j} = \sum_1^K w_k \cdot s_{hr,k,j} \tag{5}$$

or in matrix form

$$I_{hr} = W \cdot S_{hr} \tag{6}$$

$P_{hr} = \{p_{hr,j}, \ j = 1, \ldots, J\}$ – high resolution panchromatic image. It is an input for pansharpening.

2.1.3. Aim of pansharpening

The aim of pansharpening is from a given low resolution multispectral image (MS image usually exhibiting several bands) $S_{lr}$ and high resolution panchromatic image (Pan image) $P_{hr}$ to reconstruct/recover a high resolution multispectral image $S_{hr}$. In the following sections, it is shown how the preprocessing of input imagery for pansharpening can be performed in order to enhance the quality of popular CS and HPF based methods.

*2.2. Input image adjustment*

Due various reasons, e.g. calibration inaccuracies, image acquisition by different sensors, input images for pansharpening: low resolution multispectral image (or calculated intensity image) and high resolution panchromatic image may differ in their physical properties, e.g. radiances or reflectances depending on the processing level. For physically justified fusion, the same objects/classes in both images must exhibit very similar intensity/panchromatic values or more generally similar statistics. Similarity definition will depend on the particular application.

Thus, adjustment of Pan image to MS image or calculated intensity image is needed for a successful fusion. Usually histogram matching is applied. In this paper a new model-based image adjustment method is proposed. Sequential combination of both approaches is possible.

Calculation of intensity image from MS bands is treated in sub-sections 2.3.1, 2.3.2. Data provider initial weights maybe necessary as an input.

2.2.1. Histogram matching

In the workflow of most pansharpening methods two possibilities are available: Pan image histogram matching before fusion or MS image histogram matching after fusion.

By histogram matching in general we understand adjustment of histogram or more precisely its shape of one image to the histogram of another image by using cumulative histograms. We call it further a full histogram matching. An alternative approach is to match only the first-order statistics (mean and variance) of both images, as e.g. proposed in [17]. Further this type of histogram matching will be called a simple histogram matching. It will work successful only for normal distributions.

Physically justified histogram matching is reasonable between Pan and intensity images, and original MS bands and MS bands after fusion. It should be noted, that both histogram matching approaches will work for imagery of different scales/resolutions. For intensity calculation data provider weights $W_0$ are required.

2.2.2. Full histogram matching

In full histogram matching a histogram of high-resolution panchromatic image $P_{hr}$ is adjusted to the histogram of intensity image $I_{lr}$ calculated from original low resolution MS image $S_{lr}$. Of course, implicitly it is assumed, that the histogram form/shape does not depend on the scale/resolution. This image-based correction does not always leads to sufficient improvement as it is shown in the experimental section. It can be noted here, that it is possible to match Pan image to intensity image $\tilde{I}_{hr}$ calculated from up-sampled MS image $\tilde{S}_{hr}$, as it is proposed in [17]. First option, using original MS image can be preferred against interpolated/up-sampled MS image due to possible artefacts introduced by interpolation method used.

2.2.3. Simple histogram matching

Simple histogram matching denotes "the matching of the first-order statistics (mean and variance) of the Pan image to those of each MS band, or of a combination of them e.g. intensity" [17]. It should be noted, that physically it makes no sense matching Pan image to separate MS bands, thus this option is not considered in this paper.

Formulae for Pan image matching to intensity calculated from original MS bands looks like [17]

$$\hat{P}_{hr} = (P_{hr} - \mu_{P_{hr}}) \cdot \left(\sigma_{I_{lr}} / \sigma_{P_{hr}}\right) + \mu_{I_{lr}} \qquad (7)$$

in which μ and σ denote the mean and standard deviation respectively.

As already stated in the previous sub-section two options are possible: Pan adjustment to low resolution $I_{lr}$ (proposed in this paper) or Pan adjustment to high resolution $\tilde{I}_{hr}$ [17].

Further in this paper a new model-based image adjustment method is proposed.

*2.3. Model-based adjustment*

A new model-based image adjustment method performs correction of Pan image respecting physics of image formation/acquisition process, thus being superior against the image-based histogram matching.

2.3.1. Image formation/acquisition model

The idea of input image adjustment will be described below. Let's assume two optical sensors with a spectral sensor $S$ usually exhibiting several narrow spectral response functions (SRF) for each band and a panchromatic sensor $P$ with usually broad SRF, e.g. see Figure 1. It is obvious, that a simple averaging of spectral bands will be different radiometrically from a panchromatic band.

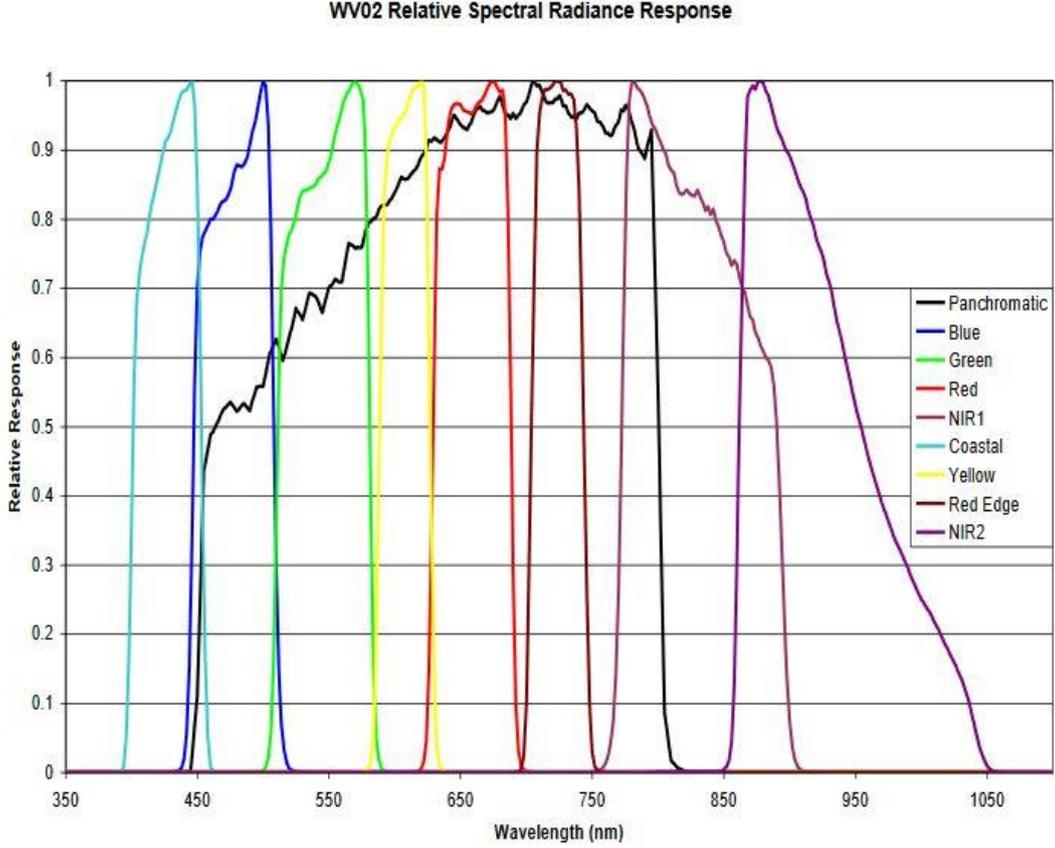

**Figure 1.** Spectral Response of the WorldView-2 panchromatic and multispectral imagery [23].

For a successful fusion of data from such two sensors the energy balance between radiances/reflectances of both sensors should hold, that is

$$\sum_1^K w_k \cdot s_{k,i} + v_i = p_i \qquad (8)$$

or in vector/matrix form

$$W \cdot S + V = P \qquad (9)$$

where $V$ is a virtual band introduced to compensate for total energy differences in different sensors and $W$ are weights for individual spectral bands. It is assumed, that weights $W$ describe properties of a multispectral signal and thus are independent on the image resolution or scale. Further, it should be noted, that a virtual band has been already introduced in [24], but without any practical solution how to estimate it.

2.3.2. Estimation of weights in low resolution image domain

Weights $\widehat{W}$ in (9) are estimated using available multispectral image in low resolution from the following equation

$$W \cdot S_{lr} = P_{lr} \qquad (10)$$

e.g. by using bounded-variable least-squares (BVLS) minimization with a constraint $0 \leq w_k \leq 1$ [25].

This optimization or weights estimation is independent on its initialization values. Thus, when the provider weights are not known, then equal values can be assumed e.g. $W_0 = 1/K$.

Here it should be noted, that other ways of estimating these weights can be used as e.g. proposed in [21] by using a linear regression algorithm. No constraint on weights can result in negative weight values, what contradicts a physical image acquisition model.

2.3.3. Estimation of virtual band in high resolution image domain

From (8) it follows that

$$v_{lr,i} = p_{lr,i} - \sum_1^K \widehat{w}_k \cdot s_{lr,k,i} \qquad (11)$$

or in vector form

$$V_{lr} = P_{lr} - \widehat{W} \cdot S_{lr} . \tag{12}$$

Further, this virtual band $V_{lr}$ is interpolated, e.g. using bicubic convolution, to a high resolution image $\tilde{V}_{hr} = \{v_{hr,j}\}$ .

Finally, in order (9) to be hold, $P_{hr}$ is corrected/adjusted using the following equation:

$$\hat{P}_{hr} = P_{hr} - \tilde{V}_{hr} . \tag{13}$$

This new panchromatic image $\hat{P}_{hr}$ ensures more similar statistical radiometric properties with $I_{hr}$ and thus can be used to enhance any pansharpening method as shown in the following section.

*2.4. Enhancement of pansharpening methods*

Below, it is shown how some of most popular pansharpening methods can be enhanced using previously introduced image adjustment methodology.

2.4.1. Enhanced CS based methods

For CS based methods using additive model the image fusion equation will look like

$$\hat{S}_{hr} = \tilde{S}_{hr} + \hat{P}_{hr} - \tilde{I}_{hr} \tag{14}$$

where $\tilde{S}_{hr}$ is an interpolated, e.g. using bicubic convolution, version of $S_{lr}$ , $\tilde{I}_{hr}$ intensity image calculated from multispectral image $\tilde{S}_{hr}$ using weights $\widehat{W}$. For multiplicative model

$$\hat{S}_{hr} = \tilde{S}_{hr} + \hat{P}_{hr} / \tilde{I}_{hr} \tag{15}$$

2.4.2. Enhanced HPF based methods

For MRA or HPF (high pass filter) based methods using additive model the equation is the following

$$\hat{S}_{hr} = \tilde{S}_{hr} + \hat{P}_{hr} - \hat{P}_{lr} \tag{16}$$

where $\tilde{S}_{hr}$ is an interpolated, e.g. using bicubic convolution, version of $S_{lr}$ , $\hat{P}_{lr}$ is a low pass filtered version of $\hat{P}_{hr}$ . For multiplicative model

$$\hat{S}_{hr} = \tilde{S}_{hr} + \hat{P}_{hr} / \hat{P}_{lr} \tag{17}$$

**3. Results**

In this section nominal and enhanced versions of CS and HPF based methods are compared on real satellite remote sensing data.

*3.1. Data*

WorldView-2 satellite remote sensing data over the Munich city in South Germany were used in the following experiments. For summary of scene details see Table 1.

Table 1. Scene parameters for WorldView-2 data over Munich city.

| Parameter | Value |
| --- | --- |
| Image date | 12-Jul-2010 |
| Image time (local) | 10:30:17 |
| Mode | PAN+MS |
| Look angle | 5.2° Left |
| Product | L2A |
| Resolution PAN (m) | 0.5 |
| Resolution MS (m) | 2.0 |
| Provider $W_0$ | [0.0074, 0.1106, 0.1787, 0.12076, 0.1987, 0.1363, 0.0959, 0.0002793] |

Thus, original multispectral image exhibits 2 m resolution and panchromatic image - 0.5 m resolution. Image size has been set to 256x256 pixels in the experiments because the relative quality performance of the methods does not depend on the image size. Data provider (DigitalGlobe) weights of spectral response functions $W_0$ are normalized to the total energy of panchromatic band (Figure 1).

*3.2. Experiment setup*

In order to have reference images for the following experiments the proposed approach presented in the publication [26] is adapted and accordingly the resolution of original data is reduced. Thus, appropriate low pass filtering as e.g. described in [27, 28] is applied on the original multispectral image and then it is sub-sampled to the resolution of 4 m. For original panchromatic image another appropriate low pass filtering [28] is applied and then it is sub-sampled to the resolution of 2 m. In Table 2 the resolution of images used in the following experiments is summarized (except Sect. 3.6.2, where original images are used, see Table 1).

**Table 2.** Resolution summary of input and reference images.

| Image | Resolution (m) |
|---|---|
| Reference $S_{hr}$ | 2 |
| Input $S_{lr}$ | 4 |
| Input/Reference $P_{hr}$ | 2 |

For visual interpretation original images are presented in Figure 2 and in a reduced scale in Figure 3.

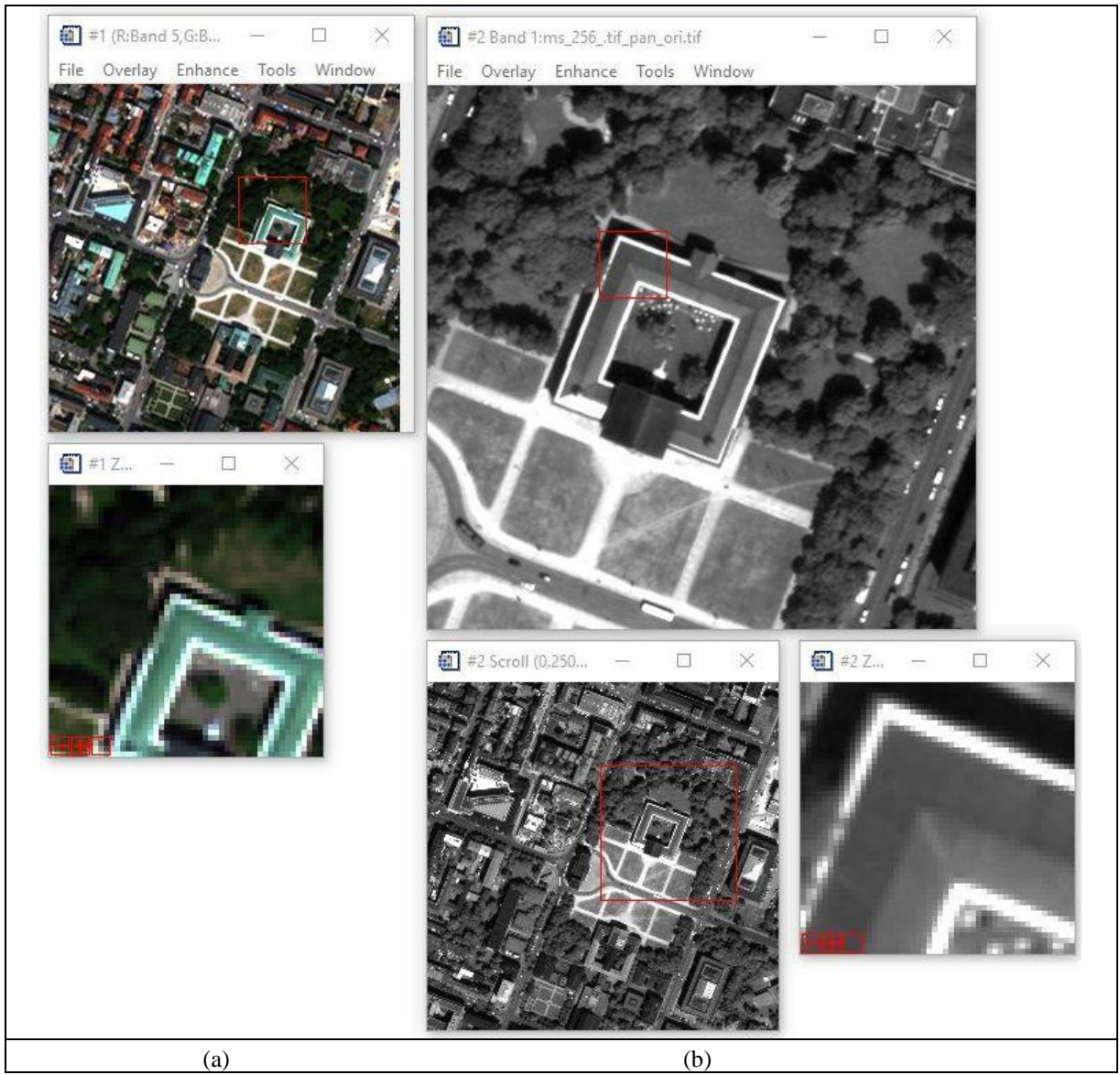

|(a)|(b)|

**Figure 2.** Multispectral image (bands: 5, 3, 2, size: 256x256) in original resolution of 2 m (a, top) and red box zoom by factor 4 (a, bottom). Panchromatic image (b, size: 1024x1024) in original resolution of 0.5 m (b, top) and scroll by factor 0.25 (b, bottom left) and zoom by factor 4 (b, bottom right).

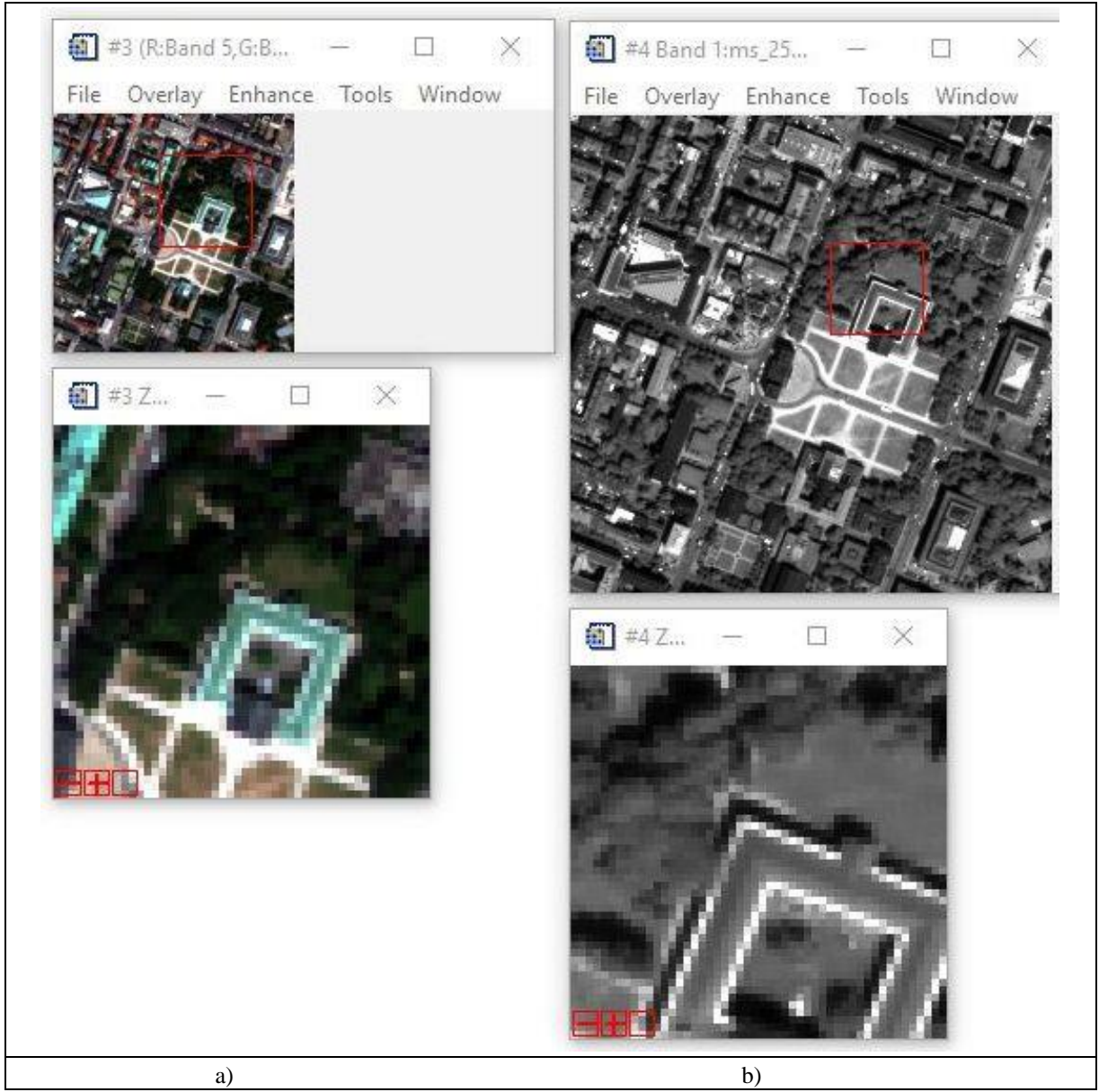

**Figure 3.** Multispectral image (bands: 5, 3, 2, size: 128x128) after low pass filtering in reduced scale 4 m (a, top) and red box zoom by factor 4 (a, bottom) and panchromatic image (b, size: 256x256) after low pass filtering in reduced scale 2 m (b, top) and red box zoom by factor 4 (b, bottom).

*3.3. Quality measure*

There exist a lot of quality measures, see e.g. overviews in [3, 29-33]. In case when the reference is known, then most measures are based on squared difference. Thus, for the relative comparison of methods it is enough the usage of one measure. When the reference image is not available (what is a nominal case in pansharpening), then available/existing quality measures e.g. such as Quality with No Reference (QNR) [34] are still not satisfactory, thus they are not considered in this paper.

In reduced resolution case the reference is known, thus the root mean squared error (RMSE) is used as a quality measure between pansharpened and reference multispectral images for each band

$$RMSE(S_{hr,k}, S_{hr,k}^{ref}) = \sqrt{1/J \cdot \sum_{1}^{J} \left(s_{hr,k,j} - s_{hr,k,j}^{ref}\right)^2} \tag{18}$$

or it's mean

$$RMSE(S_{hr}, S_{hr}^{ref}) = 1/K \cdot \sum_{1}^{K} RMSE(S_{hr,k}, S_{hr,k}^{ref}) \qquad (19)$$

Similarly $RMSE(S_{lr}, S_{lr}^{ref})$ for a low resolution multispectral images or further for panchromatic/intensity images $RMSE(P_{hr}, P_{hr}^{ref})$ can be defined.

*3.4. Interpolation of low resolution multispectral image to high resolution*

Here the influence of a simple interpolation/up-sampling of a low resolution multispectral image to a high resolution is analyzed. Any proposed pansharpening method should be better than this simple interpolation method.

Let $\tilde{S}_{hr}$ be interpolated version of original $S_{lr}$ using bicubic convolution, $\tilde{I}_{hr}$ intensity image (MSI) calculated from multispectral image $\tilde{S}_{hr}$ using data provider weights $W_0$ (Table 1) according to (5, 6). Then the values for $RMSE(\tilde{S}_{hr}, S_{hr})$ (multispectral bands) and $RMSE(\tilde{I}_{hr}, P_{hr})$ (panchromatic band) are presented in Table 3.

**Table 3.** Comparison of $RMSE(\tilde{S}_{hr}, S_{hr})$ (multispectral bands) and $RMSE(\tilde{I}_{hr}, P_{hr})$ (panchromatic band) for multispectral image interpolation using bicubic convolution.

| Band | Method MSI |
|---|---|
| 1 | 20.66 |
| 2 | 21.57 |
| 3 | 38.78 |
| 4 | 55.71 |
| 5 | 46.68 |
| 6 | 59.09 |
| 7 | 77.88 |
| 8 | 66.85 |
| Mean of all bands | **48.40** |
| Panchromatic | **37.64** |

General trend can be observed, that bands with lower numbers are better reconstructed than bands with higher numbers. It is obvious, that any correction on panchromatic band will not have impact on $RMSE(\tilde{S}_{hr}, S_{hr})$.

*3.5. Panchromatic band correction*

Panchromatic band correction influence on input data used in image fusion is analyzed in this section. In particular, it is measured how close panchromatic band can be transformed/corrected to the intensity image calculated from MS bands. In Table 4. *RMSE* values between intensity and panchromatic images in both scales are presented for various cases: no correction of panchromatic band, full Pan histogram matching (PHM, full) and simple Pan histogram matching (PHM, simple) of panchromatic band to intensity image, model-based panchromatic band correction (PC) and finally both corrections sequentially combined. PHM is applied in two versions: using intensity image in low resolution and high resolution as described in Sect. 2.2.2.

**Table 4.** Influence of Pan histogram matching and panchromatic band correction on *RMSE*.

| Correction | $RMSE(I_{hr}, P_{hr})$ | $RMSE(I_{lr}, P_{lr})$ |
|---|---|---|
| Before correction | 32.69 | 24.05 |
| PHM, full, low | 27.02 | 16.93 |
| PHM, full high | 27.59 | 17.77 |
| PHM, simple, low | 27.44 | 17.66 |
| PHM, simple, high | 26.95 | 16.89 |
| PC [1] | 22.33 | 14.81 |
| PHM, full, low + PC | 19.20 | 13.08 |
| PHM, simple, high + PC | **18.83** | **12.66** |

[1] It should be noted that for PC the same results for both initial weight sets (data provider and equal) are achieved, because it is independent on calculated intensity and thus initial weights! Moreover, it should be noted that the model-based panchromatic band correction includes update of initial weights $W_0$ which in its row has impact on the calculation of intensity images.

Both Pan histogram matching methods applied in both resolutions deliver similar results (*RMSE* values), but full histogram matching is more general (see Sect. 2.2.2), thus preferred further. The results are significantly (more than 5%) better than before correction.

Pan correction outperforms Pan histogram matching by about 5% and in total about 10% when compared with results before correction.

For sequential combined corrections: Pan histogram matching and Pan correction the performance gets further better by about 3%. Thus, for both corrections applied sequentially the total performance increase is about 14%.

*RMSE* values for low resolution (last column in Table 4) are presented for informative reasons and show quite good correlation with *RMSE* values in high resolution.

For illustration purposes the example of the virtual band in high resolution is presented in Figure 4. Additionally, an example of estimated weights is presented. For the case of sequential combination of Pan histogram matching and Pan correction the resultant/estimated values for weights are: $\widehat{W}$ = [0.180855, 0.0, 0.199144, 0.00496692, 0.305368, 0.0190767, 0.0742123, 0.0595428].

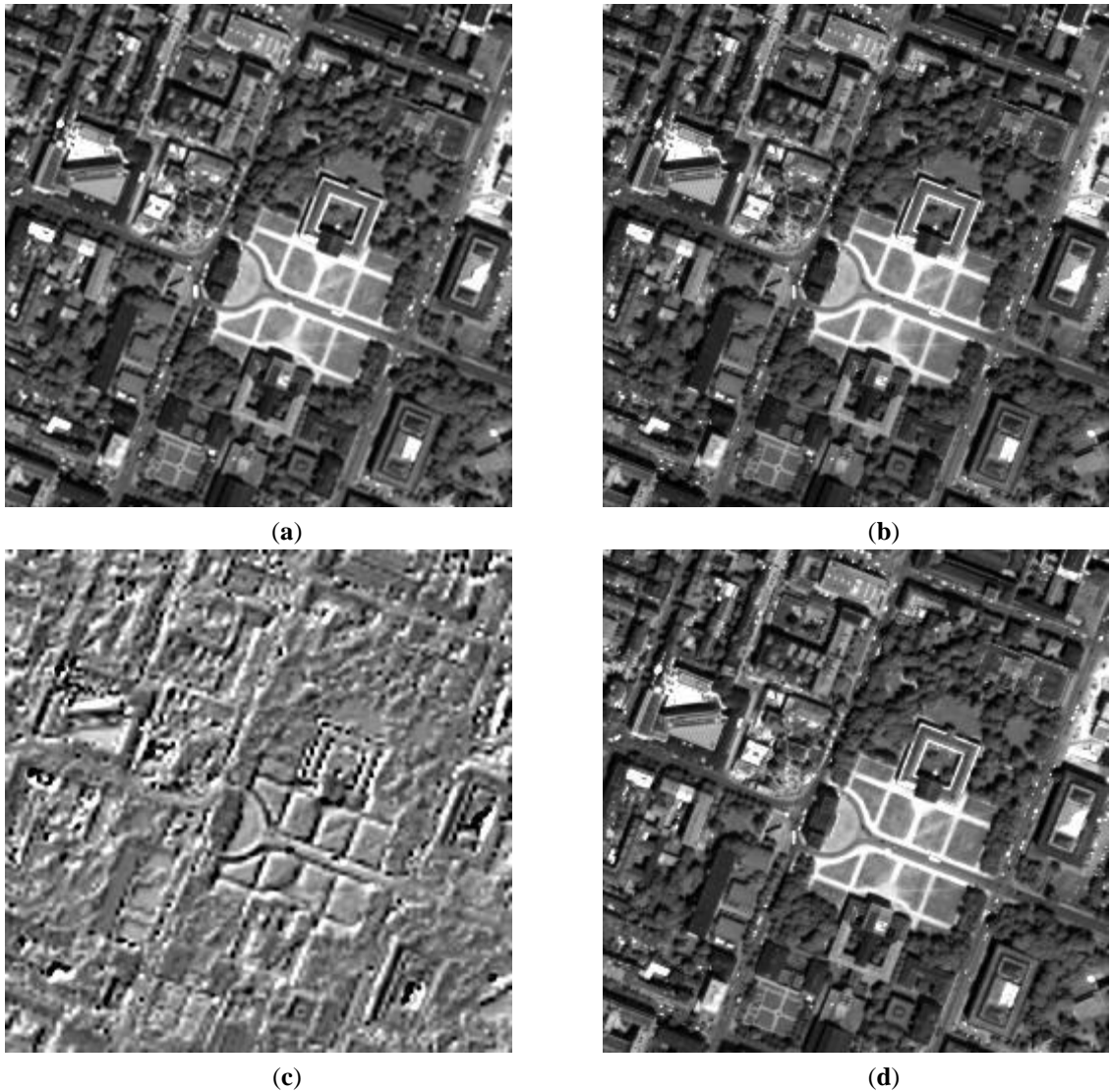

**Figure 4.** Intensity image (a), panchromatic image before correction (b), virtual band (c) and panchromatic image after correction (d).

*3.6. Comparison of pansharpening methods*

It has been shown in the previous section that the panchromatic image can be transformed/corrected much closer (in the sense of *RMSE* value) to the intensity image than the original panchromatic image is. In this section it is shown how much such transformation/correction can increase the performance of a pansharpening method. Four pansharpening methods have been compared: CS a – Component substitution with additive model [28], CS m - Component substitution with multiplicative model [28] and HPF - High pass filtering using Butterworth filter [28] in both model versions (HPF a, HPF m). For the reference purpose results for MSI – multispectral interpolation using bicubic convolution are included.

The general workflow of pansharpening methods mentioned above is:
1. Up-sample original low resolution MS image to high resolution.
2. [Histogram matching of original high resolution panchromatic image to intensity in low resolution or to intensity interpolated to high resolution. Data provider weights are needed as input.]
3. [Pan correction including weights estimation in low resolution.]
4. [Weights estimation in high resolution.]
5. Image fusion using (14-17).
6. [Histogram matching of fused MS in high resolution to original MS in low resolution.]

In [] marked are optional/enhancement steps which influence on pansharpening quality is investigated in the following section.

### 3.6.1. Reduced resolution

First, a detail investigation is performed for one of pansharpening methods (in this case the best one, see Table 6) for various parameter settings. The results (mean values of *RMSE*) are presented in Table 5 for CS multiplicative model pansharpening method depending on various input parameters: a way of correction of high resolution panchromatic image and estimation of weights $W$ used in the calculation of high resolution intensity image.

**Table 5.** $RMSE(\hat{S}_{hr}, S_{hr})$ for CS multiplicative model pansharpening method depending on input parameters: way of correction of high resolution panchromatic image and estimation of weights $W$ used in calculation of high resolution intensity image. Weights used: provider $W_0$, estimated in low resolution $\widehat{W}_{low}$ (Pan correction), estimated in high resolution $\widehat{W}_{high}$ [21]. Notation: "–" means that the correction is not applied, "+" correction applied. Upper *RMSE* value is for a case without MS histogram matching (see Sect. 2.2.1), lower – with MS histogram matching.

| Pan histogram matching | Pan correction | Weights $W$ | $RMSE(\hat{S}_{hr}, S_{hr})$ |
|---|---|---|---|
| - | - | $W_0$ | 47.64 |
|   |   |   | 40.34 |
| - | - | $\widehat{W}_{low}$ | 40.50 |
|   |   |   | 40.14 |
| - | - | $\widehat{W}_{high}$ | 40.67 |
|   |   |   | 40.66 |
| - | + | $W_0$ | 45.65 |
|   |   |   | 37.33 |
| - | + | $\widehat{W}_{low}$ | 37.61 |
|   |   |   | **36.91** |
| - | + | $\widehat{W}_{high}$ | 38.16 |
|   |   |   | 37.81 |
| Full, low | - | $W_0$ | 41.05 |
|   |   |   | 40.52 |
| Full, low | - | $\widehat{W}_{low}$ | 40.18 |
|   |   |   | 40.07 |
| Full, low | - | $\widehat{W}_{high}$ | 40.59 |
|   |   |   | 40.58 |
| Full, low | + | $W_0$ | 38.14 |
| | | | 37.57 |
| Full, low | + | $\widehat{W}_{low}$ | 36.78 |
| | | | **36.57** |
| Full, low | + | $\widehat{W}_{high}$ | 37.47 |
| | | | 37.40 |
| Full, high | - | $W_0$ | 41.52 |
| | | | 40.68 |
| Full, high | - | $\widehat{W}_{low}$ | 40.19 |
| | | | 40.09 |
| Full, high | - | $\widehat{W}_{high}$ | 40.64 |
| | | | 40.58 |
| Full, high | + | $W_0$ | 38.55 |
| | | | 37.75 |
| Full, high | + | $\widehat{W}_{low}$ | 36.65 |
| | | | **36.53** |
| Full, high | + | $\widehat{W}_{high}$ | 37.37 |
| | | | 37.34 |
| Simple, low | - | $W_0$ | 40.98 |
| | | | 40.59 |
| Simple, low | - | $\widehat{W}_{low}$ | 40.03 |
| | | | 40.00 |
| Simple, low | - | $\widehat{W}_{high}$ | 40.43 |
| | | | 40.51 |
| Simple, low | + | $W_0$ | 38.15 |
| | | | 37.74 |
| Simple, low | + | $\widehat{W}_{low}$ | 36.76 |
| | | | **36.55** |
| Simple, low | + | $\widehat{W}_{high}$ | 37.44 |
| | | | 37.36 |
| Simple, high | - | $W_0$ | 41.39 |
| | | | 40.74 |
| Simple, high | - | $\widehat{W}_{low}$ | 39.95 |
| | | | 39.98 |
| Simple, high | - | $\widehat{W}_{high}$ | 40.40 |
| | | | 40.48 |
| Simple, high | + | $W_0$ | 38.53 |
| | | | 37.82 |
| Simple, high | + | $\widehat{W}_{low}$ | 36.61 |
| | | | **36.48** |
| Simple, high | + | $\widehat{W}_{high}$ | 37.31 |
| | | | 37.28 |

From the analysis of Table 5 following main conclusions can be drawn.

MS matching after fusion practically always leads to better results and thus is always applied in further experiments.

Weights estimated in low resolution $\widehat{W}_{low}$ (result of Pan correction method) also seem to deliver better results and thus is always applied in further experiments.

All types of Pan histogram correction are quite similar and the performance is comparable with that of no or before correction. Thus, alone Pan histogram matching brings no significant performance increase.

Pan correction seems to deliver the most significant performance increase.

Sequential combination of Pan histogram matching and Pan correction leads to slightly better results when compared to alone Pan correction.

The results obtained for CS m have suggested the reduced or preferred set of parameters which have been used to compare other methods. Mean values of *RMSE* are presented in Table 6 for various pansharpening methods using selected modes of corrections.

**Table 6.** Comparison of $RMSE(\hat{S}_{hr}, S_{hr})$ for various pansharpening methods using different corrections of pan image: pan correction (PC), pan histogram matching (PHM), multispectral histogram matching (MHM).[1]

| Mode \ Method | MSI | CS a | CS m | HPF a | HPF m |
|---|---|---|---|---|---|
| Before correction | 48.40 | 45.53 | 47.64 | 43.62 | **43.35** |
| Before correction + MHM | - | 40.48 | **40.34** | 43.74 | 43.48 |
| PC + $\widehat{W}_{low}$ | - | 37.30 | **36.91** | 42.48 | 41.97 |
| PHM, full, low + PC + $\widehat{W}_{low}$ | - | 37.06 | **36.57** | 42.55 | 42.01 |
| PHM, full, high + PC + $\widehat{W}_{low}$ | - | 37.08 | **36.53** | 42.58 | 42.03 |
| PHM, simple, low + PC + $\widehat{W}_{low}$ | - | 37.05 | **36.55** | 42.52 | 41.96 |
| PHM, simple, high + PC + $\widehat{W}_{low}$ | - | 37.05 | **36.48** | 42.54 | 41.96 |

[1] It should be noted, that for all pansharpening methods (except before correction and MSI) the histogram of the resultant high resolution multispectral image is matched to the histogram of the input low resolution multispectral image. For a particular mode correspondingly corrected panchromatic image is used. Weights used are estimated in low resolution $\widehat{W}_{low}$ (Pan correction). Moreover, no corrections are applicable for MSI.

From the analysis of Table 6 following conclusions can be drawn.

Before any correction HPF methods are significantly better than CS methods. Further, application of MHM increases the performance of CS methods significantly. CS methods even outperform HPF methods significantly in this case.

PC with $\widehat{W}_{low}$ further increases the performance of all methods. Again, CS methods are significantly better than HPF and CS m is the best of all. Further, the sequential combination of both corrections brings practically no significant improvement.

Results for each band separately for illustration/informative purpose are presented in Table 7.

**Table 7.** Comparison of $RMSE(\hat{S}_{hr}, S_{hr})$ for various pan-sharpening methods for mode: Pan histogram matching + pan correction.

| Band \ Method | MSI | CS a | CS m | HPF a | HPF m |
|---|---|---|---|---|---|
| 1 | 20.66 | 19.68 | 23.93 | **19.18** | 21.18 |
| 2 | 21.57 | 19.06 | **11.56** | 19.08 | 18.38 |
| 3 | 38.78 | **25.88** | 26.18 | 32.05 | 31.73 |
| 4 | 55.71 | 37.91 | **36.37** | 47.28 | 45.84 |
| 5 | 46.68 | 30.59 | **30.28** | 38.65 | 38.41 |
| 6 | 59.09 | 43.27 | **42.01** | 51.65 | 50.52 |
| 7 | 77.88 | 64.86 | **62.27** | 71.39 | 69.56 |
| 8 | 66.85 | 55.45 | **54.08** | 61.10 | 60.06 |
| Mean of all bands | 48.40 | 37.05 | **36.48** | 42.54 | 41.96 |

Again, for most individual bands CS m delivers the best performance. Similarly, as for MSI (Table 3) the same trend can be observed, that the bands with lower numbers are better reconstructed than bands with higher numbers.

For visual interpretation purpose the results of selected pansharpening methods are presented in Figure 5.

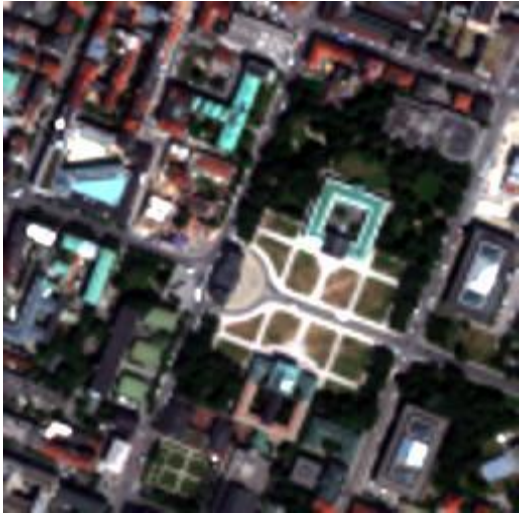 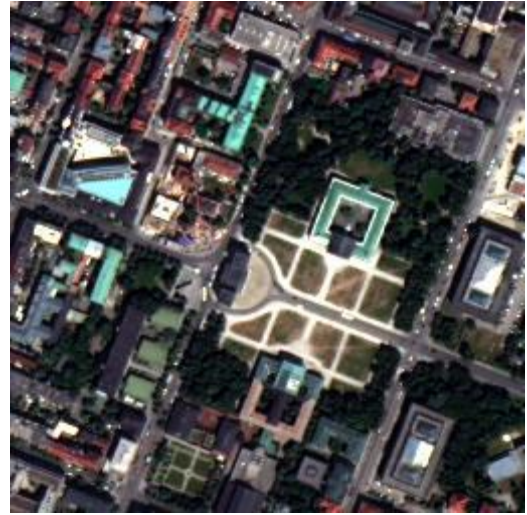

(**a**) (**b**)

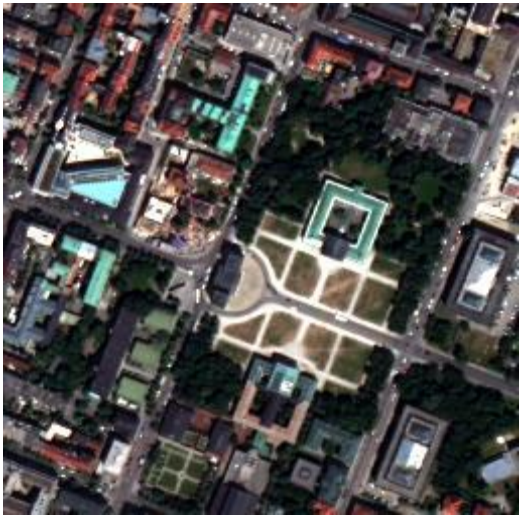 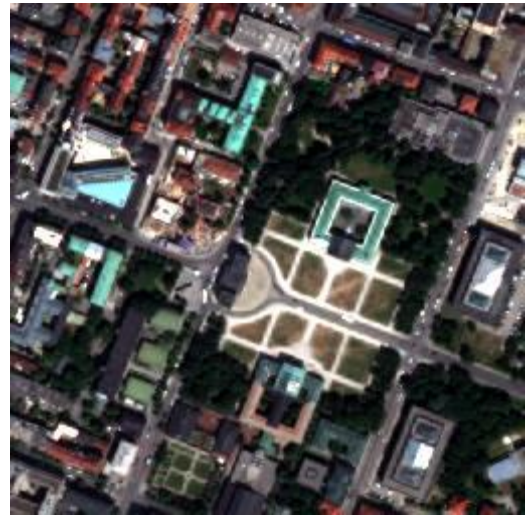

(**c**) (**d**)

**Figure 5.** Images pansharpened by methods: MSI - multispectral interpolation using bicubic convolution (a); CS a – component substitution using additive model (b); CS m – component substitution using multiplicative model (c) and HPF m - High pass filtering using Butterworth filter and multiplicative model (d) in reduced resolution (size: 256x256).

3.6.2. Original resolution

Due to the lack of a well-established quantitative quality measure in the absence of the reference only visual interpretation results of some pansharpening methods are presented in Figure 6.

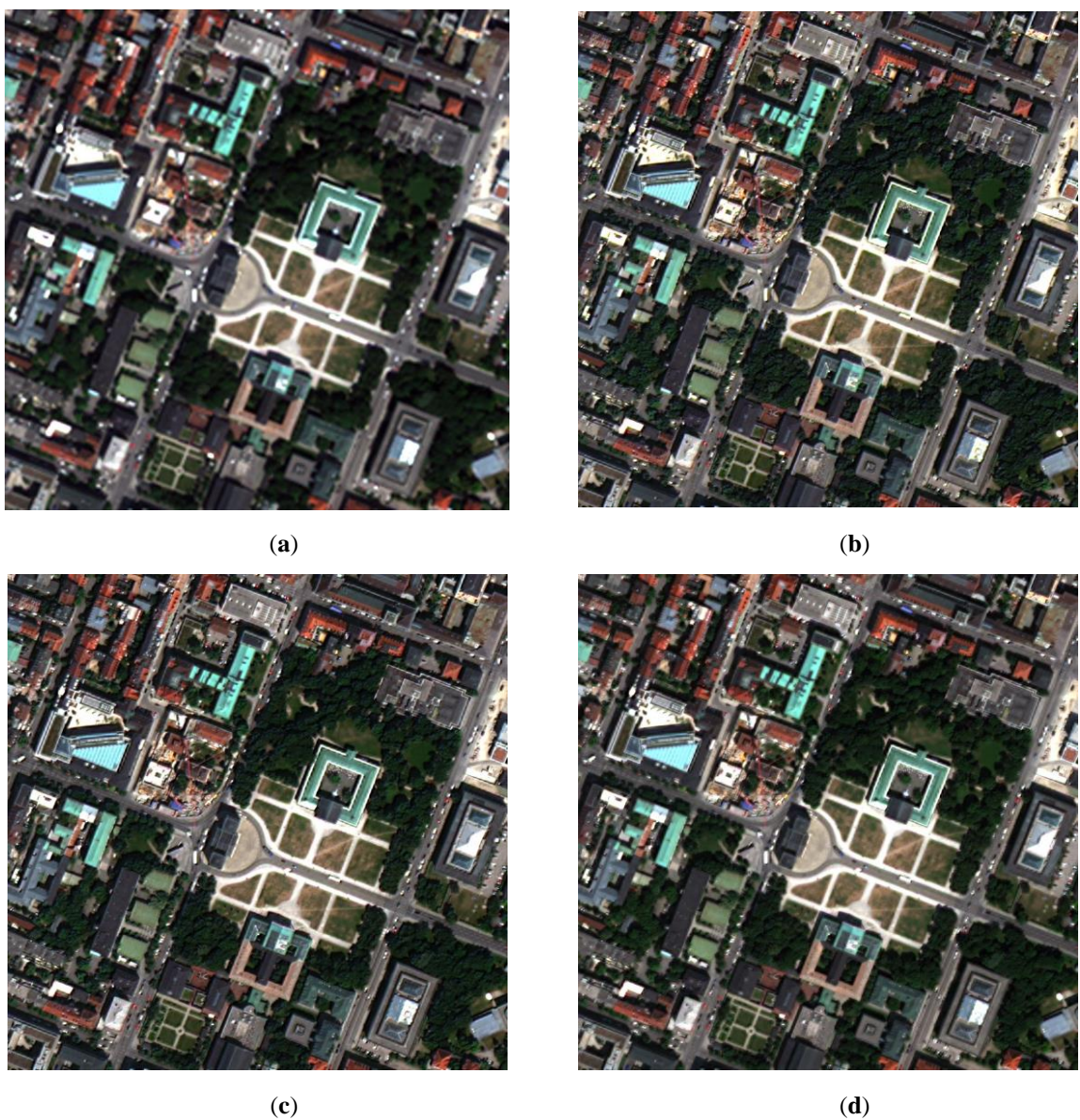

**Figure 6.** Images pansharpened by methods: (**a**) MSI - multispectral interpolation using bicubic convolution; (**b**) CS a – component substitution using additive model; (**c**) CS m – component substitution using multiplicative model and (**d**) HPF m - High pass filtering using Butterworth filter and multiplicative model in original resolution (size: 1024x1024).

Visually the results of fusion of all pansharpening methods seem to be very similar. The choice of the best method will depend on a particular application.

## 4. Discussion

Here the main results received in the previous section are discussed. Four most popular pansharpening methods: CS and HPF in both versions with additive and multiplicative model have been compared. Methods based on multiplicative model seem to outperform additive model insignificantly. Without any correction HPF methods seem to be significantly better than CS methods. Any correction: alone or in combination (Pan histogram matching, Pan correction, MS histogram matching) makes CS methods significantly better than HPF methods. Most of the performance increase is due two corrections: Pan correction and MS histogram matching. Inclusion of Pan histogram matching into the sequential combination of corrections leads to a slightly increased performance. Weights estimation as a result of Pan correction seem to be preferable in comparison to data provider initial weights. In this case data provider weights are not needed anymore and simply equal weights can be assumed.

Thus, the following general workflow of the pansharpening method can be proposed:
1. Up-sample original MS image to high resolution.
2. [Histogram matching of original high resolution panchromatic image to intensity in low resolution or to intensity interpolated to high resolution. Data provider weights are needed as input.]
3. Pan correction including weights estimation in low resolution.
4. Image fusion using CS with multiplicative model (15).
5. Histogram matching of fused MS in high resolution to original MS in low resolution.

In [] marked is an optional step of Pan histogram matching which does not has a significant influence on the pansharpening quality.

## 5. Conclusions

A model-based image adjustment to be performed for the enhancement of multi-resolution image fusion or pansharpening is proposed. Such image adjustment is needed for most pansharpening methods using panchromatic band and/or intensity image (calculated as a weighted sum of multispectral bands) as an input. For a successful fusion of data from two sensors the energy balance between radiances/reflectances of both sensors should hold. A virtual band is introduced to compensate for total energy differences in different sensors. The corrected panchromatic band is used instead of original panchromatic image in the following pansharpening. Moreover, estimated weights can be used for the intensity image calculation if appropriate. It is shown for example, that the performance quality of component substitution based methods can be increased significantly.

Presented methodology is not limited only to the pansharpening. It can be easily extended to the hyperspectral and multispectral sharpening too.

The future work could be directed towards investigation how the proposed panchromatic band correction can be used for the enhancement of the third group methods (mentioned in the introduction) such as model-based or optimization based pansharpening methods.

**Acknowledgments:** I would like to thank DigitalGlobe and European Space Imaging (EUSI) for the collection and provision of WorldView-2 scene over the Munich city.